\documentclass[reprint,aps,prd,twocolumn,superscriptaddress,nofootinbib]{revtex4-2}

\usepackage[utf8]{inputenc}
\usepackage{amsmath, amssymb, amsfonts, bm}
\usepackage{graphicx}
\usepackage[colorlinks=true, linkcolor=blue, citecolor=blue, urlcolor=blue]{hyperref}
\usepackage{microtype}
\usepackage{orcidlink}

\begin{document}

\title{Beyond FINDCHIRP: Breaking the memory wall and optimal FFTs for Gravitational-Wave Matched-Filter Searches with Ratio-Filter Dechirping}

\author{Alexander H. Nitz\orcidlink{0000-0002-1850-4587}}
\email{ahnitz@syr.edu}
\affiliation{Department of Physics, Syracuse University, Syracuse, New York 13244, USA}

\author{{Keisi Kacanja}\orcidlink{0009-0004-9167-7769}}
\affiliation{Department of Physics, Syracuse University, Syracuse, New York 13244, USA}

\author{{Kanchan Soni}\orcidlink{0000-0001-8051-7883}}
\affiliation{Department of Physics, Syracuse University, Syracuse, New York 13244, USA}

\date{\today}

\begin{abstract}
A primary bottleneck in modern FFT-based matched-filter searches for gravitational waves from compact binary coalescences is not raw processor throughput, but available memory bandwidth. Standard frequency-domain implementations, such as the FINDCHIRP algorithm, rely on streaming long template waveforms and data from main memory, which leads to significant processor stalling when template durations exceed cache capacities. In this work, we introduce \textit{Ratio-Filter Dechirping} as a solution, an algorithmic restructuring of the matched filter that transforms the operation from a memory-bound Fast Fourier Transform (FFT) into a cache-efficient, compute-bound Finite Impulse Response (FIR) convolution. By utilizing a reference template to remove common orbital phase evolution, we produce slowly changing frequency-domain ratios that can be accurately implemented as short FIR filters. This method delivers a measured speedup of $8\times$ for the core filtering loop used in offline searches and should enable $>10\times$ for low-latency analysis. We find that this approach generalizes to a variety of searches that include physical features such as finite size effects, eccentricity, and precession. By dramatically reducing the computational cost of matched filtering, this approach enables the expansion of searches into dense or high-dimensional parameter spaces, such as those for eccentric or subsolar-mass signals, that are already limited by available computing budgets. Furthermore, this framework provides a natural path for hardware acceleration on GPU architectures.
\end{abstract}

\maketitle

\section{Introduction}

The observational era of gravitational-wave (GW) astronomy is firmly established, with the currently operating Advanced LIGO~\cite{LIGOScientific:2014pky}, Virgo~\cite{TheVirgo:2014hva}, and KAGRA~\cite{KAGRA:2013rdx} observatories having reported over 200 compact binary coalescences (CBCs) to date~\cite{LIGOScientific:2018mvr, Nitz:2021uxj, KAGRA:2021vkt, LIGOScientific:2025hdt, Venumadhav:2019lyq, Wadekar:2023gea}. These observations have provided a view of the population of stellar-mass black holes and neutron stars, offering insights into stellar evolution~\cite{Gerosa:2021mno,Edelman:2021fik,Zevin:2020gbd,LIGOScientific:2025pvj,LIGOScientific:2025brd}, the nuclear equation of state~\cite{Capano:2019eae,De:2018uhw,LIGOScientific:2018cki, Reed:2025sqh}, and fundamental tests of general relativity~\cite{LIGOScientific:2019fpa,LIGOScientific:2020tif,Wang:2021gqm,LIGOScientific:2025obp}. However, the majority of current detections represent a limited region of the available parameter space: quasi-circular, spin-aligned binaries.

A significant observational blind spot remains in the detection of signals with more complex morphologies, specifically those exhibiting high levels of eccentricity, precession, or those with subsolar-mass components, although significant efforts are made in these areas~\cite{McIsaac:2023ijd, Phukon:2024amh, Schmidt:2024hac, Kacanja:2024hme, Hanna:2024tom, Phukon:2025cky, Dhurkunde:2026jhp}. Binaries formed via dynamical capture in dense stellar environments, such as globular clusters, nuclear star clusters, or active galactic nuclei (AGN) disks, are expected to retain significant residual eccentricity at frequencies where ground-based detectors are most sensitive~\cite{Samsing:2013kua, Rodriguez:2018pss}. Evidence for such dynamical formation has already appeared in the first half of the third observing run, with events like the high-mass binary GW190521~\cite{LIGOScientific:2020iuh, Romero-Shaw:2020thy} and the neutron star-black hole candidate GW200105~\cite{LIGOScientific:2021qlt, Morras:2025xfu,Kacanja:2025kpr} exhibiting features consistent with a dynamical or triple binary origin~\cite{Romero-Shaw:2025otx}. Simultaneously, the search for subsolar-mass compact objects ($< 1 M_\odot$) provides a direct probe for new physics, including primordial black holes~\cite{Carr:2020gox} or dissipative dark matter composites~\cite{Shandera:2018xkn}, which would have profound implications for our understanding of the early universe.

Complementary to these template-based methods, loosely-modelled ``burst'' searches~\cite{Drago:2020kic} and machine-learning-based (ML) classifiers~\cite{Gebhard:2019ldz, Marx:2024wjt, Silver:2025crh} have been developed to detect short duration signals. While ML-based approaches are often cited for their low computational latency, many performance comparisons in the literature do not represent a true algorithmic speedup; instead, they often reflect the hardware disparity between highly-parallel GPU inference and suboptimal, serial CPU implementations of matched filtering~\cite{Schafer:2022dxv}. Furthermore, as demonstrated in recent mock data challenges~\cite{Schafer:2022dxv}, these methods have yet to demonstrate competitive sensitivity for long-duration signals that dominate the computational budget of modern searches.

Despite their scientific potential, comprehensive searches for these sources are often limited by computational cost. The template banks required to search for eccentric or low-mass signals grow rapidly in density~\cite{Kacanja:2024hme,Kacanja:2024pjh}. For instance, a search for subsolar-mass objects can require over 20 million templates~\cite{Nitz:2021vqh}, and even then, such searches often fail to cover the full potential parameter space. Standard search pipelines based on the FINDCHIRP~\cite{Allen:2005fk} algorithm, such as implemented in PyCBC~\cite{Usman:2015kfa} are optimized for frequency-domain matched filtering using the Fast Fourier Transform (FFT). While computationally efficient for standard signals, these pipelines face a hardware-level bottleneck known as the ``Memory Wall.'' On modern architectures, raw processor throughput (FLOPS) has dramatically outpaced memory bandwidth (DRAM access speeds). In standard FFT-based searches, long template waveforms must be streamed from main memory, leading to persistent processor stalling as CPU cores wait for data that cannot be served fast enough to satisfy the arithmetic units.

Several alternative search architectures for matched filtering have been developed that optimize for different scenarios such as low-latency analysis, but have also been applied to high throughput searches. The \texttt{gstLAL} pipeline~\cite{Messick:2016aqy} utilizes a single Value Decomposition (SVD) to project templates onto a reduced basis, significantly decreasing the number of operations required for otherwise intractable direct time-domain convolution. The \texttt{MBTA} pipeline~\cite{Adams:2015ulm} uses FFT-based matched filtering, but employs a multi-band approach, downsampling lower-frequency components to reduce the data volume, while \texttt{SPIIR}~\cite{Chu:2020pjv} implements the matched filter in the time domain using Infinite Impulse Response (IIR) filters to facilitate low-latency analysis. While most effective at producing low-latency optimized analysis, these approaches have been competitive with direct FFT-based matched filtering for high throughput searches as well.

In this work, we present a framework that addresses the Memory Wall directly: \textbf{Ratio-Filter Dechirping}. The method restructures the matched filter from a memory-bound FFT operation into a cache-efficient, compute-bound convolution. By dechirping target templates relative to a coarse grid of reference signals, we produce frequency-domain ratios that map to very short time-domain filters. The dechirping procedure is conceptually similar to the first step of heterodyning techniques also employed for parameter estimation~\cite{Cornish:2010kf,Zackay:2018qdy}, but with significant changes to application and relaxing of smooth ratio requirements. This approach restores high arithmetic intensity to the CPU, delivering a measured $8\times$ speedup for archival searches and has the potential for $>10\times$ improvement for low-latency analyses such as PyCBC Live~\cite{DalCanton:2020vpm}.

This mathematical foundation of ratio-based reconstruction has also been independently discussed in recent work by Murakami et al. (2025)~\cite{Murakami:2025rpx}, who demonstrated its utility in reducing pregenerated waveform storage footprint, but demonstrated only a modest change to performance. We demonstrate an approach that achieves substantial improvements to \textit{hardware throughput} and \textit{cache locality} aspects of the problem. We demonstrate a matched filtering algorithm that can be implemented straightforwardly into existing search architectures; this includes how to generate appropriate template banks, optimized generation of FIR ratio filters, and a demonstration of the realized performance improvements. We find that this method is also highly generalizable to a variety of search spaces that include the effects of eccentricity and precession.

\section{The Computational Bottleneck: The Memory Wall}

\begin{figure}[htbp]
    \centering
    \includegraphics[width=0.45\textwidth]{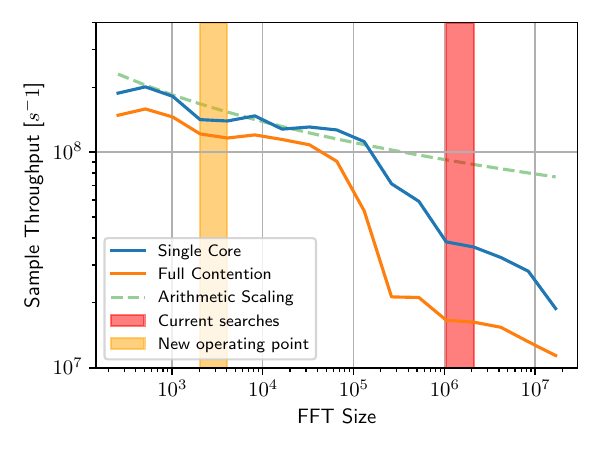}
    \caption{The sample rate throughput of the Fourier transform as a function of FFT size benchmarked on a representative cluster node (Haswell architecture) when the machine is fully utilized (orange) and when it is otherwise unoccupied (blue). The expected scaling just accounting for arithmetic operations is shown (green dotted). We see that the FFT performance scales initially with the expected computational cost; at larger sizes memory access becomes the bottleneck so the performance degrades. The current operating point of typical GW searches is shown (red) along with the target operating point of our new algorithm (yellow). An $8\times$ performance improvement is possible if the final algorithm is FFT dominated. Similar scaling is also found on a variety of x86 CPU architectures.}
    \label{fig:fft_scaling}
\end{figure}

The foundational algorithm for gravitational-wave matched-filtering is \texttt{FINDCHIRP}~\cite{Allen:2005fk}, which leverages the Fast Fourier Transform (FFT) to compute the correlation between detector data and a bank of potential signal templates. This frequency-domain approach has been the standard for decades, as it reduces the computational complexity of the naive time-domain matched filter from $O(N^2)$ to $O(N \log N)$, where $N$ is the number of samples in a data segment. However, the computational efficiency of this algorithm is limited by the memory access patterns required for large FFT sizes.

Modern search pipelines, such as those implemented in \texttt{PyCBC}~\cite{Usman:2015kfa}, typically operate in an offline, archival mode or a low-latency, real-time mode~\cite{Nitz:2018rgo}. In both cases, the primary computational cost is dominated by the inverse FFT (IFFT) required to produce the complex SNR time series. To maximize throughput, these pipelines are designed to saturate the floating-point capabilities (FLOPS) of a processor. However, as templates become longer—reaching durations of hundreds or thousands of seconds for binary neutron stars or subsolar-mass objects—the volume of data required for a single matched-filter operation grows significantly.

A modern CPU core features a deep memory hierarchy, starting with small, fast L1 and L2 caches (typically hundreds of kilobytes to a few megabytes) and extending to larger L3 caches and main system memory (DRAM). The peak arithmetic performance of the core can only be sustained if the required data is resident in the L1/L2 cache. For standard CBC search sample rates ($2048$--$4096$~Hz), the memory footprint of a frequency-domain template and the associated data vector for a signal lasting $\sim 100$~seconds is approximately $10$--$20$~MB. This footprint exceeds the typical per-core L2 cache capacity of current server-class processors, forcing the core to retrieve data from the much slower memory.

This disparity creates the ``Memory Wall'': a regime where the arithmetic units of the processor are frequently idle, stalled while waiting for data to be streamed from memory. We quantify the severity of this bottleneck in Figure~\ref{fig:fft_scaling}. Even under ideal operational conditions (shown in blue), throughput remains high for small FFT sizes that fit within the cache. However, as the FFT size increases beyond the cache limit ($\sim 10^5$ samples), throughput plummets by a factor of $5-8\times$. In production environments where all cores are occupied and competing for the shared memory bus (shown in orange), the performance degradation is most severe.

We note that all testing was done with single-core configurations. Multi-core FFTs can better use multiple cores while staying within cache constraints, however, finding optimal plans can often be difficult, and the availability of high-throughput computing resources strongly favors single-core operation in practice~\cite{Jayatilaka:2017twe,Weitzel:2017ocs}.

This bottleneck is not merely a matter of wall-clock time; if the computational cost of a search increases eight-fold due to memory stalls, the depth of the template bank must be commensurately reduced to fit within available computing budgets. This trade-off directly impacts the sensitivity of searches which require exceptionally dense banks to maintain a high minimal match.

\begin{figure*}[htbp]
    \centering
    \includegraphics[width=2\columnwidth]{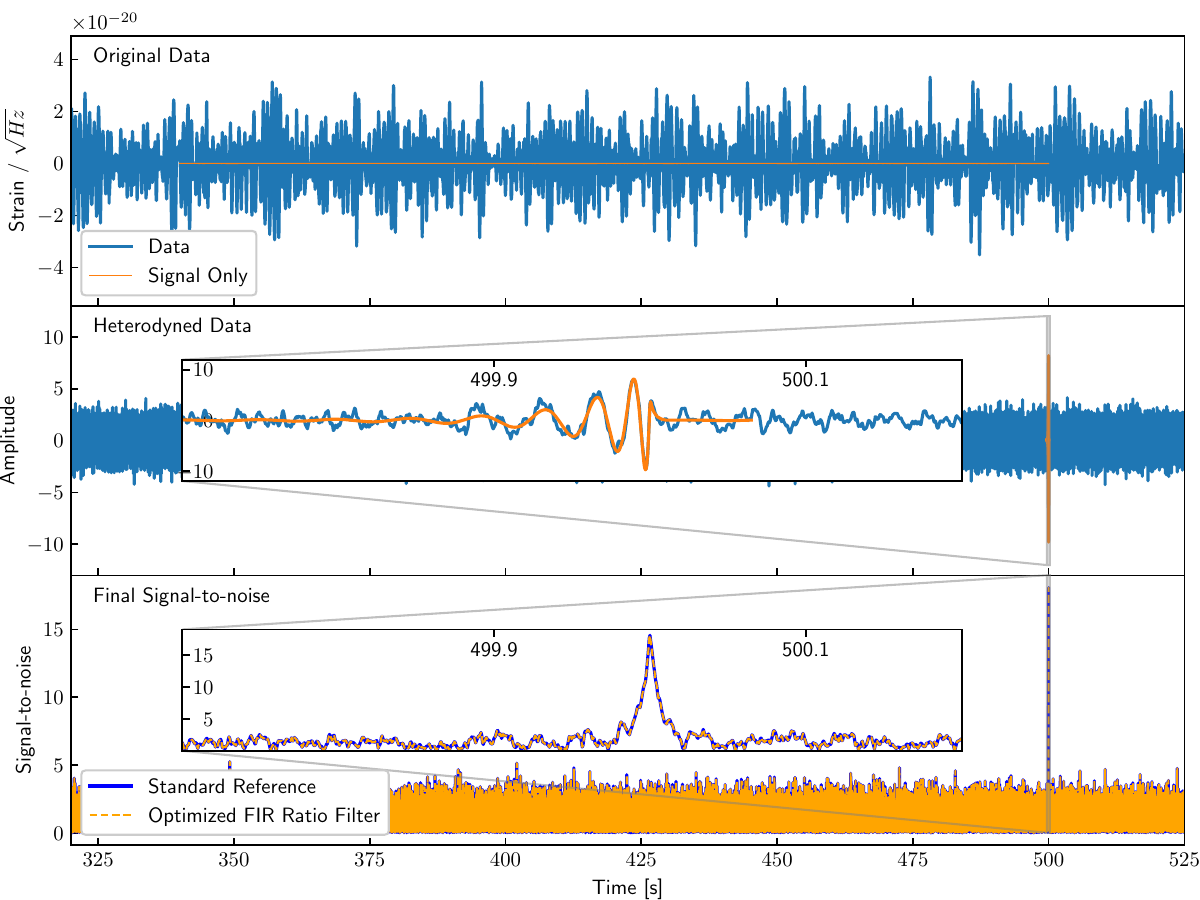}
    \caption{Schematic representation of the two-step Ratio-Filter Dechirping process. The detector data (top) is first matched-filtered against a coarse reference template to produce a reference SNR time series (middle). This latter step can be thought of as a combined filtering stage which both overwhitens and heterodynes the data. For visual clarity, only the real part of the otherwise complex filter output is shown. This time series is then convolved with a short FIR ratio filter to reconstruct the final SNR for a specific target template (bottom). The contribution to the data of an embedded simulated binary neutron star signal is highlighted in orange in both the top and middle panels. The original BNS signal which is $\sim 170s$ in duration is efficiently compressed to a region of only a fraction of a second after the heterodyning stage. As can be seen, the final SNR result is nearly identical to the original reference SNR calculated using a single-stage matched filter. The improvement in computational performance occurs because the application of the FIR ratio filter can be done separately on much shorter segments of data (e.g. O(1s) as opposed to O(100s).}
    \label{fig:demo}
\end{figure*}

\section{Ratio-Filter Dechirping Theory}

The detection of a gravitational-wave signal $h(t)$ in a data stream $s(t)$ relies on the optimal statistic for colored stationary Gaussian noise: the matched filter. The core operation of a search pipeline is the calculation of the complex Signal-to-Noise Ratio (SNR) time series, $(s|h)(t)$. In the frequency domain, this is defined as:
\begin{equation}
(s|h)(t) = 4 \int_{0}^{\infty} \frac{\tilde{s}(f) \tilde{h}^{*}(f)}{S_n(f)} e^{2\pi i f t} df\,,
\label{eq:mf_standard}
\end{equation}
where $\tilde{s}(f)$ is the Fourier transform of the detector strain, $\tilde{h}(f)$ is the frequency-domain template, and $S_n(f)$ is the one-sided noise Power Spectral Density (PSD).

Standard algorithms such as \texttt{FINDCHIRP}~\cite{Allen:2005fk} implement Eq.~\ref{eq:mf_standard} by performing a point-wise multiplication of the data and the conjugate template in the frequency domain, followed by an inverse Fast Fourier Transform (IFFT). This approach is highly efficient because the FFT algorithm reduces the complexity of searching over all possible arrival times $t$ from $O(N^2)$ to $O(N \log N)$. However, as established in Section II, this efficiency is only fully realized when the data vectors remain within the processor's cache.

Our proposed method, \textit{Ratio-Filter Dechirping}, maintains the underlying optimality of the matched filter while restructuring the calculation to preserve cache locality. This restructuring relies on the mathematical associativity of convolutions along with the fact that searches typically use dense set of template waveforms e.g. a template bank containing signals with some commonality. We begin by introducing a reference template $\tilde{h}_{\text{ref}}(f)$—typically a signal from a comparatively coarse template bank that is nearby in parameter space to the target template $\tilde{h}(f)$. We then express the target template as:
\begin{equation}
\tilde{h}(f) = \tilde{h}_{\text{ref}}(f) \cdot \left( \frac{\tilde{h}(f)}{\tilde{h}_{\text{ref}}(f)} \right)\,.
\label{eq:ratio_def}
\end{equation}
Substituting Eq.~\ref{eq:ratio_def} into Eq.~\ref{eq:mf_standard}, we can re-write the SNR time series as:
\begin{equation}
\begin{split}
(s|h)(t) &= \mathcal{F}^{-1} \left[ \frac{\tilde{h}(f)}{\tilde{h}_{\text{ref}}(f)} \cdot \left( 4 \frac{\tilde{s}(f) \tilde{h}_{\text{ref}}^{*}(f)}{S_n(f)} \right) \right] \\
         &= \mathcal{F}^{-1}\left[ \frac{\tilde{h}(f)}{\tilde{h}_{\text{ref}}(f)} \right] * (s|h)_{\text{ref}}(t)\,,
\end{split}
\label{eq:dechirp_theory}
\end{equation}
where $(s|h)_{\text{ref}}(t)$ is the pre-calculated SNR time series for the reference template.

The physical motivation for this approach is the concept of ``dechirping.'' The reference SNR time series $(s|h)_{\text{ref}}(t)$ captures the bulk of the rapidly oscillating orbital phase evolution common to the local region of the parameter space. The ratio $R(f) = \tilde{h}(f) / \tilde{h}_{\text{ref}}(f)$ therefore removes this dominant chirp behavior, resulting in a slowly varying function in the frequency domain. By the properties of the Fourier transform, a slowly varying frequency-domain signal corresponds to a highly localized kernel in the time domain.

The resulting time-domain representation, $r(t) = \mathcal{F}^{-1}[R(f)]$, is a very short Finite Impulse Response (FIR) filter, often requiring only $O(10^{2-3})$ taps (see Sec. \ref{filter_construction}). This is orders of magnitude shorter than the original template duration. By transforming the matched filter into a convolution with a short FIR filter, we shift the computational load from streaming long waveforms to performing small, dense convolutions that fit entirely within the L1/L2 cache. These convolutions can still be performed in the frequency-domain to preserve the logarithmic scaling advantages.

The act of filtering data then follows a two-step procedure as outlined in Fig.~\ref{fig:demo}. Filtering a long stream of data involves splitting it into overlapping blocks, e.g. the overlap-add method of fourier-domain convolution. Because the matched-filter output is corrupted for the duration of the template, to ensure sufficient valid output at high performance, these blocks are typically chosen to be a multiple of the template duration itself. Although the full data must still be analyzed, the performance advantage is the result of being able to analyze the data in shorter chunks.

This architecture offers a unique scaling advantage over other established matched-filtering approaches. In Table~\ref{tab:method_comparison}, we compare the algorithmic complexity and the typical floating-point operation (FLOP) multipliers for a standard binary neutron star search.

\begin{table}[htbp]
    \centering
    \caption{Algorithmic complexity comparison for generating $N$ output SNR samples using various matched filtering methods. We show only the leading order cost. $T$ is the template length, $R$ is SVD rank, $M$ is number of frequency bands, $C$ is IIR filter order, and $K$ is the overlap-add block size. All the algorithms have a trivial scaling with the number of samples N; the ``Multiplier'' estimates the total FLOP count per sample for a typical BNS search. In addition to the cache locality advantages, our proposed algorithm also reduces the required number of floating point calculations. For the ratio-filter method there is an additional cost for filtering the coarse templates. However, given that these are a small fraction of the total effective bank size (e.g. $<1\%$) the contribution is negligible.}
    \label{tab:method_comparison}
    \begin{tabular}{l l r}
        \hline
        \textbf{Method} & \textbf{Scaling Cost} & \textbf{Multiplier} \\
        \hline
        Standard FFT & $O(N \log_2 N_{\text{block}})$ & $\sim 20$ \\
        GstLAL (SVD)~\cite{Messick:2016aqy} & $O(N \cdot R)$ & $100 - 500$ \\
        SPIIR (IIR)~\cite{Chu:2020pjv} & $O(N \cdot C)$ & $100 - 200$ \\
        MBTA~\cite{Adams:2015ulm} & $O(\sum N_i \log T_i) + O(N \cdot M)$ & $\sim 15$ \\
        \hline
        \textbf{Ratio-Filter} & $\mathbf{O(N \log_2 K)}$ & $\mathbf{\sim 11}$ \\
        \hline
    \end{tabular}
\end{table}

While Standard FFTs require massive block sizes ($N_{\text{block}} \approx 2^{20}$) to amortize the overhead of long signals, the Ratio-Filter's compact kernel allows for much smaller processing blocks ($K \approx 2048$). This reduces the scaling multiplier from $\sim 20$ to $\sim 11$, offering a reduction in the operation count of $2\times$, while simultaneously solving the Memory Wall bottleneck; this compounds the overall performance improvement.

\section{Construction of the Search Bank}

To utilize the Ratio-Filter Dechirping framework, one can use a two-stage hierarchical template bank. Standard gravitational-wave (GW) search banks are designed to ensure that any signal within the targeted parameter space has a normalized overlap, or ``match,'' with at least one template above a specific minimal threshold, typically $0.965$ or $0.97$~\cite{Usman:2015kfa}. Our approach augments this search bank with a coarse ``reference'' bank that provides the anchor points for the dechirping operation.

As a proof of concept, we generated two template banks covering a portion of the binary neutron star (BNS) parameter space using the Advanced LIGO O3 power spectral density for chirp masses $\mathcal{M} \in [1.2, 1.3] M_\odot$. The coarse reference bank was generated using stochastic placement tools~\cite{Kacanja:2024pjh} with a minimal match threshold of $\sim 0.5$. The dense target bank was generated with the standard search threshold of $0.965$. In this configuration, the number of templates in the reference bank is $<1\%$ of the size of the full search bank. Consequently, the initial computational cost of performing standard matched filtering for the reference signals is negligible compared to the total search budget. We also note that more extreme ratios are observed in dense parameter spaces such as when additional physical parameters are added.

\begin{figure}[htbp]
    \centering
    \includegraphics[width=0.45\textwidth]{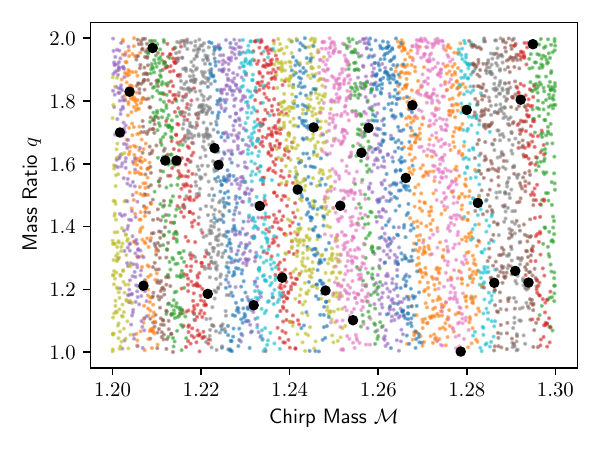}
    \caption{Visualization of the hierarchical template bank structure in the component mass parameter space for a slice of the BNS region ($\mathcal{M} \in [1.2, 1.3] M_\odot$). The large black points represent the coarse reference bank (minimal match $\sim 0.5$), while the dense colored points represent the standard search bank (minimal match $0.965$). The color of each target template indicates its association with the nearest reference neighbor, ensuring the resulting ratio waveform corresponds to a short, computationally efficient FIR filter.}
    \label{fig:bank_twostep}
\end{figure}

The logical association between target templates and their corresponding references can be precalculated during the bank generation phase. Each template in the dense bank is mapped to the reference template that maximizes the match. This association logic is critical to the performance of the algorithm; if a target template is associated with a reference that is too distant in parameter space, the phase differences will oscillate too rapidly in the frequency domain, leading to a time-domain FIR filter that is too long to fit in the processor's L1/L2 cache. By maintaining a coarse match of $\sim 0.5$, we ensure that the physical parameters of the target and reference signals remain sufficiently close to guarantee a compact FIR kernel.

\section{Robust Design of FIR Ratio Templates}
\label{filter_construction}
The efficiency of the Ratio-Filter framework depends on the ability to represent the frequency-domain ratio $R(f) = \tilde{h}(f) / \tilde{h}_{\text{ref}}(f)$ as a compact time-domain kernel $r(t)$. While the mathematical associativity of convolutions guarantees that an exact inverse Fourier transform of the ratio exists, a naive implementation via windowing and decimation often results in excessively long filters due to spectral leakage and sharp frequency cutoffs. To ensure the FIR filters are as short as possible while maintaining high signal fidelity, we employ a directly optimized and self-validating strategy.

\subsection{Filter Optimization via $\chi^2$ Minimization}

Rather than utilizing standard windowing techniques, we precalculate the FIR templates by minimizing the $\chi^2$ difference between the frequency-domain response of the discrete FIR taps and the target ratio $R(f)$. The optimization problem is framed as finding a vector of time-domain taps $\mathbf{a}$ such that the resulting frequency response $\mathbf{A} = \mathcal{F}(\mathbf{a})$ minimizes:
\begin{equation}
\chi^2 = \sum_{i} w(i) \left| R(f_i) - \sum_{k} a_k e^{-2\pi i f_i t_k} \right|^2\,,
\end{equation}
where $w(i)$ is a frequency-dependent weight, often chosen to be proportional to the signal power spectral density to prioritize accuracy in the detector's most sensitive bands. This approach is particularly robust because it accounts for the explicit lower and upper frequency cutoffs ($f_{\text{low}}, f_{\text{high}}$) inherent in gravitational-wave data. Because the effective bandpassing of the data is already performed by the reference template, the ratio filter only needs to accurately model the relative phase and amplitude evolution within that bandwidth.

To prevent numerical error accumulation artifacts, we incorporate a regularization ``ridge'' into the filter construction~\cite{Hoerl:1970}. By adding a penalty term proportional to the magnitude of the taps (L2 regularization), we ensure that the resulting filters are smooth and well-behaved, even in regions where the target signal power is low or undefined (e.g. outside of some frequency cutoffs). While mathematically rigorous, the numerical construction could otherwise result in filters which oscillate wildly with extreme magnitudes outside of the strictly controlled frequency regions, resulting in unbounded numerical error accumulation for the frequency band of interest.

\begin{figure}[htbp]
    \centering
    \includegraphics[width=\linewidth]{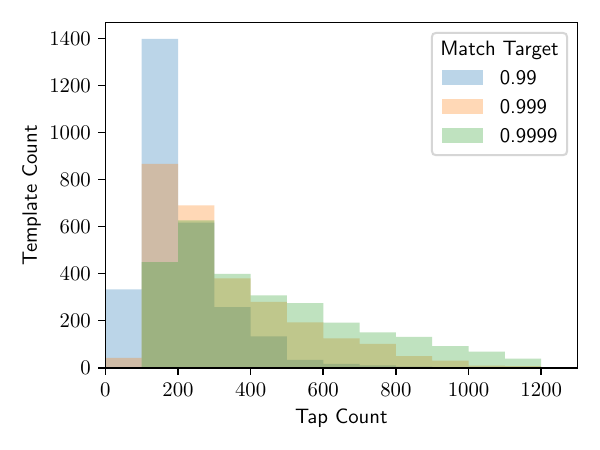}
    \caption{The distribution of FIR ratio filter size in terms of the number of taps required for a target match to the original target signal of 0.99 (blue), 0.999 (orange), or 0.9999 (green) for a representative BNS template bank. The full template bank is designed for a minimal match of 0.965, while the coarse bank is chosen so that it is $\sim150\times$ smaller in size than the full bank. The waveform is sampled at a rate of 2048 Hz. The short tap size means that small blocks can be used in a standard overlap-add matched filtering analyses, e.g. $2^{11-12}$ points or 1-2s, while still ensuring that the majority of output is uncorrupted by filter wraparound. As expected, the tap size distribution shifts to larger values with increasing match target, however, we observe that it only grows logarithmically with the mismatch, making it feasible to reach high accuracy where required.}
    \label{fig:taps}
\end{figure}

We dynamically tune the number of taps to achieve a target match accuracy with the original target template. This scheme is analogous to the ``compressed waveform'' approach currently utilized in \texttt{PyCBC} to reduce the cost of template generation~\cite{pycbc-github}. As seen in Fig.~\ref{fig:taps}, for our example BNS scenario, the distribution of taps is centered on $\sim 200$ to achieve an accuracy of $99\%$; this can be increased to an average of $\sim 500$ to achieve $99.99\%$ match.

\subsection{Morphological Generalization}

\begin{figure}[htbp]
    \centering
    \includegraphics[width=\linewidth]{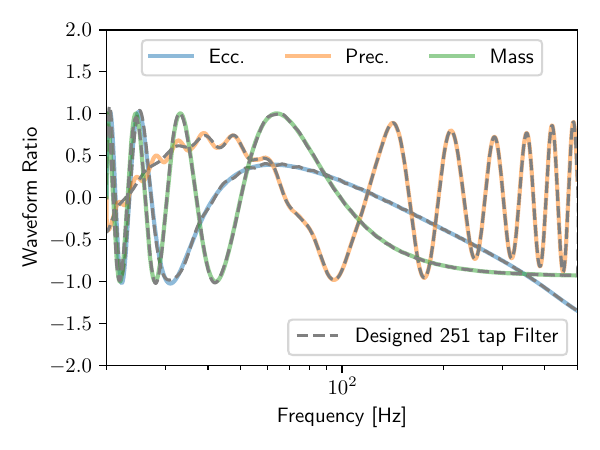}
    \caption{Frequency-domain representation of the ratio $\tilde{h}(f)/\tilde{h}_{\text{ref}}(f)$ for three distinct physical deviations from a common reference template (a $5.4 M_\odot - 1.4 M_\odot$ NSBH system). The real part of the ratio is shown. The target waveforms introduce a shift in chirp mass (green), orbital eccentricity (blue), and spin precession (orange). Despite the match between the reference and target being $<0.6$, a designed 251-tap FIR filter (dashed grey lines) achieves a match of $>0.999$ to the target signal.}
    \label{fig:waveform_ratio}
\end{figure}

A critical requirement for this method is that a single reference template must be capable of dechirping a wide variety of target templates, even if they represent significantly different physical parameters. In Fig.~\ref{fig:waveform_ratio}, we demonstrate this capability by analyzing the dominant $(2,2)$ mode of a $5.4 M_\odot - 1.4 M_\odot$ neutron star-black hole (NSBH) reference system.

We examine three distinct deviations: a shift in chirp mass, the introduction of orbital eccentricity, and the addition of spin-precession. In each of these cases, the initial match between the reference and the target signal is low ($<0.6$). However, the frequency-domain ratio remains a slowly varying function. As shown, a single 251-tap FIR filter (corresponding to a duration of $\sim 0.15\,\text{s}$) is sufficient to capture these complex morphological changes with an reconstructed match of $>0.999$. This robustness allows a sparse coarse bank to effectively anchor a vast parameter space.

While the current discussion focuses on the dominant $(2,2)$ quadrupole mode, the Ratio-Filter framework is extensible to higher-order harmonics ($l,m$ modes). These can be handled either by direct inclusion in the template bank, where the ratio remains smooth for nearby inclinations~\cite{Harry:2017weg}, or through mode-by-mode filtering where each harmonic is treated with its own ratio filter against a reference mode~\cite{Capano:2016dsf}. Because the cost of applying the FIR filter is so low, even searches requiring multiple modes per template remain computationally efficient compared to the standard full-FFT approach.

\section{Performance Validation}

To ensure that the theoretical advantages of Ratio-Filter Dechirping translate into practical speedups, it is necessary to optimize the entire search pipeline. While the core filtering operation is the primary target, auxiliary steps—such as data preparation, normalization, and signal consistency tests—must also scale effectively or remain computationally negligible to prevent them from becoming new bottlenecks.

\subsection{Profiling the Filtering Loop}

To validate the efficiency of the FIR-based approach, we conducted a benchmark profiling of a prototype filtering loop. The benchmark assumes a batch-processing mode (100 templates per batch) with a standard FIR filter size of 251 taps. The results are summarized in Table~\ref{tab:filter_benchmark}.

\begin{table}[htbp]
    \centering
    \begin{tabular}{l r r}
        \hline
        \textbf{Operation} & \textbf{Time (s)} & \textbf{Load (\%)} \\
        \hline
        Data Copy   & 0.0147 & 0.7\% \\
        Data FFT    & 0.0587 & 2.8\% \\
        Vector Multiply & 0.1307 & 6.3\% \\
        IFFT        & 1.6652 & 80.3\% \\
        Peak Find   & 0.1971 & 9.5\% \\
        Trigger Accumulation & 0.0081 & 0.4\% \\
        \hline
        \textbf{Loop Total} & \textbf{2.0746} & \textbf{100\%} \\
        \hline
    \end{tabular}
    \caption{Benchmark profiling of the core filtering loop for a batch of 100 templates using 251-tap FIR filters. Despite the significant overall speedup compared to standard FFT searches, the IFFT remains the dominant operation. However, because the block size is small enough to remain cache-resident, the absolute time is reduced by nearly an order of magnitude.}
    \label{tab:filter_benchmark}
\end{table}

The profile confirms that even with the proposed method, the IFFT of applying the now short-duration FIR ratio templates remains the dominant cost, accounting for approximately $80\%$ of the loop time. However, because the Ratio-Filter method utilizes much smaller block sizes ($K \approx 2048$) than a standard search ($N \approx 2^{20}$), these IFFTs are executed entirely within the L1/L2 cache. This cache residency prevents the processor stalling observed in Section II. Furthermore, auxiliary operations like vector multiplication and SNR peak identification scale alongside the convolution, ensuring that they do not become new bottlenecks. We find that the overall performance scaling to standard filtering matches the expectation from Fig.~\ref{fig:fft_scaling}, achieving an $8\times$ improvement in this core filtering loop.

\subsection{Optimized SNR Normalization}

Properly normalizing the SNR time series requires calculating the template variance $(h|h)$. In standard pipelines, this involves a frequency-domain integral over the full bandwidth, which must be updated as the detector PSD fluctuates. This step can be optimized by exploiting the short duration of the ratio filter through Parseval's theorem.

Starting with the frequency-domain definition and substituting $\tilde{h}(f) = \tilde{h}_{\text{ref}}(f) \cdot R(f)$, we have:
\begin{equation}
    (h|h) = 4 \Re \int_{0}^{\infty} \frac{|\tilde{h}_{\text{ref}}(f)|^2}{S_n(f)} |R(f)|^2 df\,.
\end{equation}
Applying Parseval's theorem, we convert this into a time-domain summation:
\begin{equation}
    (h|h) = \int_{-\infty}^{\infty} (h_{\text{ref}}|h_{\text{ref}})(t) \cdot \mathcal{F}^{-1} \left[ |R(f)|^2 \right](t) \, dt\,,
\end{equation}
where $(h_{\text{ref}}|h_{\text{ref}})(t)$ is the pre-calculated autocorrelation of the reference template. Since $R(f)$ corresponds to a short FIR filter, the integral collapses into a dot product over only a few active taps. 

This calculation can be further optimized by utilizing ``drift'' measurements to account for linear PSD changes over time, rather than performing full recalculations~\cite{Zackay:2019btq, Mozzon:2020gwa}. In standard PyCBC Live operations~\cite{Nitz:2018rgo}, the normalization only needs to be updated once per $\sim 60$ matched filtering operations, making its contribution to the total computational budget negligible.

\subsection{Signal Consistency Tests}

Signal consistency tests, such as the time-frequency $\chi^2$ test~\cite{Allen:2004gu}, are vital for distinguishing astrophysical signals from non-Gaussian noise transients (glitches). In standard searches, these tests typically account for $\sim 4\%$ of the total computational budget. Schematically, this signal consistency test divides the frequency space into bands and assesses that the SNR is accumulated proportionally in each frequency band. To ensure these tests do not become a new bottleneck as matched-filtering throughput increases, the statistic can be optimized by extending the dechirping logic to these frequency bins. Specifically, the sub-SNR time series for a target template can be reconstructed by applying the ratio-filter kernel to the pre-calculated sub-SNRs of its associated reference template. This approach avoids the redundant cost of re-filtering the raw detector data for each of the frequency bins, allowing the consistency test to scale with the same hardware efficiency as the primary matched filter.

 However, as there are a wide variety of noise-mitigation measures used in practice~\cite{Allen:2004gu,Schmidt:2024kxy,Davies:2020tsx,Joshi:2025nty}, new heuristics and optimizations may be needed for some searches to prevent noise mitigation from becoming a dominant cost once the matched-filtering cost reductions are realized in production analyses.

\section{Impact on Search Architectures}

The transition from a memory-bound FFT architecture to a compute-bound FIR convolution framework has implications for the design of next-generation gravitational-wave search pipelines. Beyond the raw throughput gains on CPU-based clusters, the Ratio-Filter framework enables more efficient low-latency alerts and provides a natural path forward for hardware acceleration.

\subsection{Low-latency Analysis and Early Warning}

Low-latency pipelines, such as \texttt{PyCBC Live}~\cite{Nitz:2018rgo, DalCanton:2020vpm}, currently face a significant computational penalty compared to archival analyses. In the standard overlap-save method used for real-time processing, generating the next short block of SNR data (e.g., 1-8 seconds) requires re-processing data for the entire duration of each template in the bank (often $> 100$ seconds). This results in partially redundant computation.

The Ratio-Filter method eliminates this redundancy. Because the ratio filter is a compact time-domain FIR kernel (typically $< 0.25$ seconds), calculating the next second of SNR requires only the most recent reference SNR data plus a small buffer for filter settling. This architecture brings FFT-based searches into parity with time-domain optimized pipelines regarding latency, while maintaining the superior scaling of the FFT transform-based method. This capability is essential for pre-merger detection and early-warning alerts for binary neutron star coalescences, where every second of lead time is vital for multi-messenger follow-up~\cite{Sachdev:2020lfd, LIGOScientific:2017ync}.

\subsection{Suitability for Hardware Acceleration}

The Ratio-Filter framework fundamentally alters the arithmetic intensity of the matched filter, making it an ideal candidate for acceleration on Graphics Processing Units (GPUs) and other SIMD (Single Instruction, Multiple Data) architectures. Modern GPUs are capable of massive floating-point throughput but are often throttled by global memory latency. Standard long-template FFTs are difficult to optimize on GPUs because they require large memory allocations that do not fit in the fast, on-chip shared memory, leading to significant overhead from global memory access. In contrast, the Ratio-Filter method replaces these large-scale transforms with short, dense convolutions. By increasing the number of calculations performed per byte of data loaded, the Ratio-Filter method enables significantly higher device utilization and throughput compared to traditional frequency-domain filtering.

\subsection{Enabling Stochastic and Hybrid Pipelines}

The efficiency of the Ratio-Filter engine also complements emerging search strategies that move beyond fixed template grids. For example, stochastic search algorithms such as particle swarm optimization~\cite{Pal:2023dyg} can benefit from the ability to rapidly explore local parameter spaces around candidate points using cheap FIR perturbations rather than expensive full-waveform generations. Furthermore, the compact nature of the FIR ratio provides a natural bridge to hybrid Machine Learning (ML) pipelines~\cite{Gebhard:2019ldz, Schafer:2021cml}. Because the ratio-filter operation is structurally similar to the small-kernel convolutions used in deep learning, it can serve as a lightweight, physics-based validation layer to confirm candidates identified by neural network classifiers—or to reject glitches—at speeds compatible with GPU-accelerated inference rates.

\section{Discussion and Conclusion}

In this work, we have presented Ratio-Filter Dechirping, a framework designed to bypass the ``Memory Wall'' that currently limits the sensitivity of gravitational-wave matched-filter searches. By restructuring the filtering operation from a memory-bound FFT process into a cache-efficient FIR convolution, we have demonstrated a path to significantly increase the throughput of compact binary coalescence pipelines.

Our profiling results in Section VI confirm that this approach delivers a nearly order-of-magnitude increase in filtering throughput. By reducing the effective processing block size from $N \approx 2^{20}$ to $K \approx 2048$, we ensure that the active data residency remains within the CPU's L1/L2 caches. This is particularly critical in real-world production environments.

The robustness of the $\chi^2$-optimized FIR templates allows for a sparse reference bank—requiring only $\sim 1\%$ of the total template count—to anchor a dense search for signals with complex morphologies. We have shown that even significant physical deviations in chirp mass, spin-precession, and orbital eccentricity can be captured with high fidelity (match $>0.999$) using compact 251-tap filters. 

Looking forward, the Ratio-Filter framework provides a natural architecture for hardware acceleration. Its reliance on small, dense convolutions makes it ideally suited for the shared-memory architectures of modern GPUs, where global memory latency is a persistent bottleneck. Furthermore, the elimination of the re-processing penalty in low-latency pipelines offers a clear advantage for early-warning systems, providing vital lead time for multi-messenger follow-up in future observing runs.

As we transition into the era of third-generation detectors such as the Einstein Telescope~\cite{Punturo:2010zz} and Cosmic Explorer~\cite{2019arXiv190704833R,Evans:2023euw}, the signals we seek will grow longer and the template banks will grow larger. Algorithmic innovations that account for the underlying hardware constraints of high-performance computing will be a prerequisite for the next generation of discovery. Ratio-Filter Dechirping provides a scalable, physics-informed solution to these challenges, ensuring that our search capabilities keep pace with the increasing sensitivity of the global detector network.

\acknowledgements
AHN, KK, and KS acknowledge support from the NSF grant PHY-2309240. We acknowledge the support of Syracuse University for providing the computational resources through the OrangeGrid High Throughput Computing (HTC) cluster supported by the NSF award ACI-1341006.

\bibliography{main}

\end{document}